\documentstyle[prb,aps,floats,epsf]{revtex}
\begin{document}
\twocolumn[
\hsize\textwidth\columnwidth\hsize\csname@twocolumnfalse\endcsname
\draft    
\title{Doping Dependence of the Pseudogap in  
La$_{\bf {2-x}}$Sr$_{\bf {x}}$CuO$_{\bf {4}}$}
\author{J. G. Naeini, X. K. Chen, and J. C. Irwin}
\address{Department of Physics, Simon Fraser University, Burnaby, 
British Columbia, V5A 1S6, Canada}
\author{M. Okuya,$^{\ast}$ T. Kimura,$^{\dagger}$ and K. Kishio}
\address{Department of Superconductivity, University of Tokyo, 
Bunkyo-ku, Tokyo 113, Japan}
\date{Phys. Rev. B59, 1 April 1999}
\maketitle
\begin{abstract} 
We report the results of Raman scattering experiments on single 
crystals of La$_{\rm {2-x}}$Sr$_{{\rm x}}$CuO$_{4}$ that span the range  
from underdoped (x = 0.10) to overdoped (x =0.22). 
The spectra are consistent with the existence of a strong anisotropic 
quasiparticle interaction that results in a normal
state depletion of spectral weight from regions of the Fermi
surface located near the zone axes.  The strength of the interaction 
decreases rapidly with increasing hole concentration and the spectral 
evidence for the pseudogap vanishes when the optimum doping level is reached.
The results suggest that the pseudogap and superconducting gap arise from 
different mechanisms. 
\end{abstract} 

]

\section{Introduction}
It has been recognized for some time \cite{allen,bozovic}  
that the normal state electronic properties of the high temperature 
superconductors (HTS) 
are very different from those of a conventional Fermi liquid. It is also 
generally agreed that an understanding of the normal state properties is  
important in identifying the physical mechanism 
responsible for superconductivity in HTS.  
As a result an increasing amount of attention has focused on 
the temperature region T$\,>\,$T$_c$  and recent studies 
\cite{randeria,puchkov,tatiana,slichter,ino,marshall,harris,ding,norman2,loram1,chen1,hackl,blumberg,quilty,tallon1} 
of underdoped compounds have revealed remarkable deviations from Fermi 
liquid behavior. These results 
\cite{puchkov,tatiana,slichter,ino,marshall,harris,ding,norman2,loram1,chen1,hackl,blumberg,quilty,tallon1} 
have been interpreted in terms of the opening of a normal state 
pseudogap (PG), a term that is generally used to mean a large suppression
of low energy spectral weight. 

Despite an extensive experimental effort the physical 
origin of the PG remains undetermined. Some of the proposed mechanisms 
involve precursor pairing \cite{randeria,emery,anderson} at T$\,>\,$T$_c$ . 
For example Emery and Kivelson \cite{emery} have suggested that 
incoherent pairs form above T$_c$, with phase coherence and hence 
superconductivity occurring at T$_c$. Or if one has separation of charge 
and spin \cite{anderson} d-wave pairing of 
spinons  can take place at T$\,>\,$T$_c$, with charge condensation at T$_c$. 
On the other hand in the nearly antiferromagnetic Fermi liquid (NAFL) model 
proposed by Pines and coworkers \cite{pines1,pines2,pines3}, 
the PG and superconducting mechanisms are competing. In this model the PG or 
spectral weight depletion (SWD) is caused by strong anisotropic 
antiferromagnetic (AFM) interactions that are peaked near the AFM ordering 
vector {\bf Q}$\,=\,(\pi,\pi)$. 

To obtain additional, complementary information on the nature and extent of 
the PG regime we have carried out electronic Raman scattering 
experiments on La$_{\rm {2-x}}$Sr$_{{\rm x}}$CuO$_{4}$ [La214(x)].  This  
compound is an excellent material for these studies for several reasons.  
It has a relatively simple structure with a single CuO$_2$ plane in the 
primitive cell. Thus the results are not influenced by
structural complications such as those introduced by the  chains
in YBa$_2$Cu$_3$O$_{\rm y}$ (Y123) or the structural modulations 
in Bi$_2$Sr$_2$CaCu$_2$O$_{\rm z}$ (Bi2212). 
Finally the hole concentration is determined simply by the Sr 
concentration if oxygen stoichiometry is maintained. 
As a result, one can obtain \cite{kishio} high quality and well characterized 
single crystals of La214 that enable one to study the systematic 
evolution of the electronic properties throughout the 
complete doping range.  We have measured the 
relative strengths and frequency distributions of the B$_{1g}$ and 
B$_{2g}$ Raman continua as a function of temperature and doping.  
The results of these experiments are consistent with a PG mechanism 
that is intrinsic to a single CuO$_2$  layer and is present only in 
the underdoped regime of La214. Our results also imply that the energy 
scale E$_g$ associated with the PG is quite different from the 
superconducting gap energy $\Delta$.

\section{Experiment}

The samples studied in this work were grown by a travelling 
floating-zone method \cite{kishio}.  The crystals used here were 
cut from larger single crystals and typically had dimensions of about 
$2 \times 2 \times 0.5$ mm$^3$.  Characterization of the samples included 
susceptibility and transport measurements of their critical temperatures 
\cite{kishio}.  The sample surfaces were prepared for light scattering 
experiments by polishing with diamond paste and etching \cite{werder} with a 
bromine-ethanol solution.  Finally the samples were oriented in the basal 
plane using Laue x-ray diffraction patterns.  The physical parameters 
characterizing the La214 samples are summarized in Table I.  

\begin{table}[htb]
\centering
\begin{tabular}{|c|c|lc|}  
La$_{\rm {2-x}}$Sr$_{{\rm x}}$CuO$_{4}$&Nominal Sr content (x)&T$_c$ 
($^{\circ}$K)&\\ \hline 
Underdoped&0.10&27&\\ \hline
Underdoped&0.13&35&\\ \hline
Optimally doped&0.17&37&\\ \hline
Overdoped&0.19&32&\\ \hline
Overdoped&0.22&30&\\  
\end{tabular}
\vspace{0.10in}
\caption{The physical parameters characterizing the samples studied in 
this paper. T$_c$ was determined magnetically and 
x is the nominal hole concentration$^{23}$.}
\label{tab:xT$_c$}
\end{table}

The Raman spectra were obtained in a quasibackscattering geometry using 
the 488.0 or 514.5 nm lines of an Ar$^+$ laser.  In most cases the 
light was focussed onto the sample using a cylindrical lens, but on occasion 
a combination of a spherical and cylindrical lenses was used to reduce the 
dimensions of the laser beam at the sample.  In all cases the incident 
power level was controlled to be less 
than 10 W/cm$^2$.  With these power levels an estimate of about 11$\,$K has 
been obtained for the amount of local heating by the laser.  
This estimation was obtained from measurements of the power dependence of 
the spectra at temperatures near T$_c$, and the Stokes/anti-Stokes ratio 
at somewhat higher temperatures.  The temperatures reported  
in this paper are the ambient value plus 11$\,$K, 
or the estimated temperature in the excited region of the sample.

In a Raman spectrum the intensity of the scattered light is proportional
to the square of the Raman tensor $\gamma$, whose components can be 
selected by appropriate choices of the incident and scattered polarizations
\cite{chen2}. In this paper spectra were obtained in the  
$z(xy)\bar{z}$ and $z(x^{\prime}y^{\prime})\bar{z}$ 
scattering geometries where the letters inside the brackets denote 
the polarizations of the incident and scattered light.
Here the $x$ and $y$ axes are parallel to the Cu-O bonds 
and $x^{\prime}$ and $y^{\prime}$ axes are rotated by 45$^{\circ}$ 
with respect to $(x,y)$.  
The xy component of the Raman tensor $\gamma_{xy}$ transforms as k$_x$k$_y$ 
or the B$_{2g}$ irreducible representation of the D$_{4h}$ point group, 
and $\gamma_{x^{\prime}y^{\prime}}$ as (k$^2_x-\,$k$^2_y$) or 
B$_{1g}$.  Thus $\gamma$(B$_{1g}$) must vanish along the diagonal directions 
($\pm 1,\pm 1$) in k-space and $\gamma$(B$_{2g}$) must vanish along the 
k$_x$ and k$_y$ axes.  These two scattering geometries allow us to probe 
complementary regions of the FS \cite{chen2} since the B$_{1g}$ channel will 
sample regions of the FS located near the 
k$_x$ and k$_y$ axes while the B$_{2g}$ channel will probe regions located 
near the diagonal directions.  

Throughout this paper we deal with the frequency distribution
of the Raman response function $\chi^{\prime\prime}(\omega,$T)
which is obtained by dividing the measured intensity by the thermal 
factor [1--exp(-$\hbar \omega$/$k_B$T)]$^{-1}$. 

\section{Results}

\begin{figure}[htb]
\centerline{\epsfxsize=2.5 in \epsffile{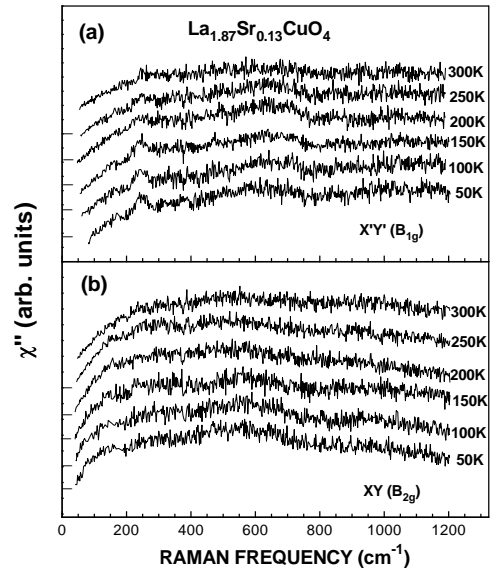}}
\vspace{0.10in}
\caption{ The B$_{1g}$ (a) and B$_{2g}$ (b) response functions 
of the underdoped La214(0.13) crystal for temperatures between 50K and 300K.
The spectra are offset vertically for clarity.}
\label{1}
\end{figure}

A detailed description of Raman spectra of La214(0.17) sample, 
which we will consider to be representative of the optimally doped state, 
has been published recently \cite{chen2}. In this paper 
we will focus on the changes that occur in the spectra of the underdoped
and overdoped compounds. We will first discuss spectra
obtained from the underdoped samples (x = 0.10 and 0.13). The normal state 
response functions are shown in Fig. 1 for the 0.13 sample and the low 
temperature response functions for both samples are presented in Fig. 2. 
As is shown in Fig. 1a, the magnitude of $\chi^{\prime\prime}$(B$_{1g}$)
is independent  of temperature, to within the accuracy of our measurements 
(15\%). From figures 2a and 2b, it is evident that the B$_{1g}$ spectra 
obtained for each sample, above and below T$_c$, are essentially identical 
and are thus unaffected by the superconducting (SC) transition. 
In direct contrast, the low energy B$_{2g}$ spectra of both 
underdoped compounds (Figs. 2c and 2d) undergo a superconductivity 
induced renormalization (SCIR) as the temperatures of the samples are lowered 
through  T$_c$.   
The observed immunity of the B$_{1g}$ spectra to passage into the 
SC state, and contrasting SCIR in the B$_{2g}$ channel, 
is analogous to the behavior observed  in underdoped Y123 \cite{chen1}.   

\begin{figure}[htb]
\centerline{\epsfxsize=2.5in \epsffile{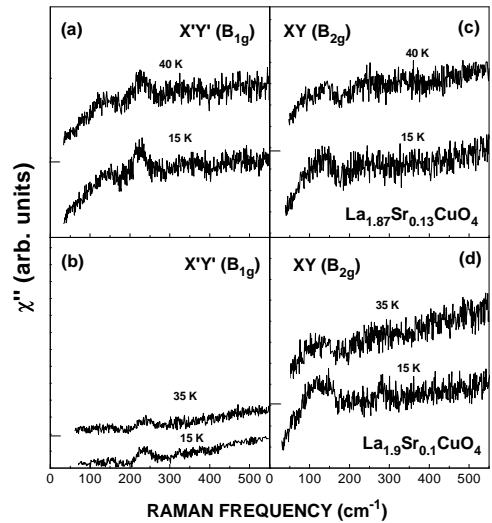}}
\vspace{0.10in}
\caption{The B$_{1g}$ response above and below T$_c$  for the La214(0.13) 
crystal (a) and the La214(0.10) crystal (b).  The B$_{2g}$ response for the 
same crystals are shown in (c) and (d).}
\label{2}
\end{figure}
\begin{figure}[htb]
\centerline{\epsfxsize=2.7 in \epsffile{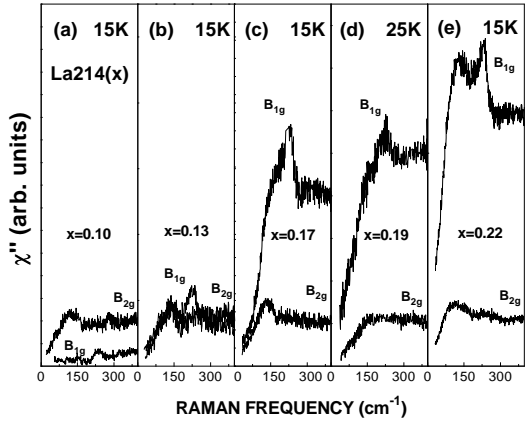}}
\vspace{0.12in}
\caption{Direct Comparison of the low energy B$_{1g}$ and 
B$_{2g}$ Raman response functions measured at temperatures
below T$_c$ in La214(x) for (a) x = 0.10; (b) x = 0.13; (c) x = 0.17; 
(d) x = 0.19; and (e) x = 0.22.  The scale of $\chi^{\prime\prime}$ is 
the same for all five frames.}
\label{3}
\end{figure}

We have also obtained spectra from two overdoped crystals (x = 0.19 
and 0.22) and we have found that the magnitude of the response functions 
$\chi^{\prime\prime}$(B$_{1g}$) in both samples 
(for $\omega \geq \,300\,cm^{-1}$) 
are again approximately independent \cite{jafar1,jafar2} of temperature.  
In contrast to the situation in the underdoped samples, there is a pronounced 
SCIR that occurs at T$_c$ in {\em both} the B$_{1g}$ and B$_{2g}$ channels
for both the optimally doped \cite{chen2} and the overdoped samples 
\cite{jafar1,jafar2}.  Finally the low temperature spectra of the overdoped 
crystals, in both  B$_{1g}$ and B$_{2g}$ channels, are compared to the 
underdoped and optimally doped spectra in Fig. 3. 

\begin{figure}[htb]
\centerline{\epsfxsize=2.4 in \epsffile{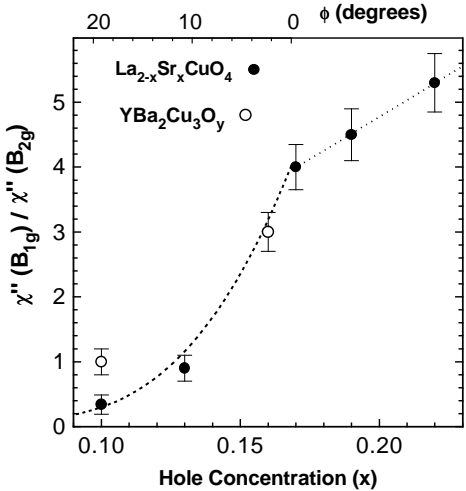}}
\vspace{0.12in}
\caption{The ratio R of the integrated Raman response functions plotted as 
a function of hole concentration for La214 ($\bullet$) and Y123 ($\circ$).  
The ratio was determined for $0 \leq \omega \leq 600\,cm^{-1}$ 
in all cases. The dashed line represents the calculated R using equation (1) 
for different values of $\phi$ (Fig. 5) superimposed using the top scale.
The dotted line serves only as a guide to the eye.}
\label{4}
\end{figure}

As a measure of the strength of the response functions we will use the 
integral of $\chi^{\prime\prime}(\omega)$ over the low frequency
spectral region.  
In this context the strength of $\chi^{\prime\prime}$(B$_{2g}$) is 
approximately the same (Fig. 3) for all the crystals studied here, 
while the strength of $\chi^{\prime\prime}$(B$_{1g}$) changes dramatically 
as we proceed  from the underdoped to the overdoped samples. 
The ratio R = $\int{\rm d}\omega\, \chi^{\prime\prime}$(B$_{1g}$) / 
$\int{\rm d}\omega \,\chi^{\prime\prime}$(B$_{2g}$), 
and therefore the strength of the B$_{1g}$ Raman response, increases 
by a factor of about 16 as we go from the underdoped x = 0.10 sample to 
the optimally doped sample with x = 0.17.  
The ratio R continues to increase, but at a slower rate as one proceeds into 
the overdoped regime. This result is summarized quantitatively 
in Fig. 4, where we have plotted the ratio R as a function of hole 
concentration. For comparison purposes, data obtained previously \cite{chen1} 
on Y123  are also included \cite{tallon2} in Fig 4. It is evident that the 
ratio R for the two compounds exhibits a similar trend with doping. 

\section{Discussion}

\subsection{Spectral Weight Depletion}

Relatively few Raman experiments have been carried out to 
investigate the nature of the PG. Nemetschek  {\em et al}. \cite{hackl} 
have attributed a small loss of spectral weight at $\omega \leq 700\,cm^{-1}$ 
in the B$_{2g}$ channel to the presence of a PG.  Blumberg {\em et al}. 
\cite{blumberg}, in spectra obtained from underdoped Bi2212 have observed 
a sharp feature at about 600$\,cm^{-1}$ which they have attributed to a bound 
state associated with precursor pair formation above T$_c$.  
In our studies of underdoped Y123 and La214 we have focused on  
the striking loss of spectral weight in the B$_{1g}$ channel, 
and the contrasting constancy of the B$_{2g}$ spectra, features which appear 
to be common \cite{chen1,katsufuji,opel} to the hole doped cuprates.  

\begin{figure}[htb]
\centerline{\epsfxsize=2.6 in \epsffile{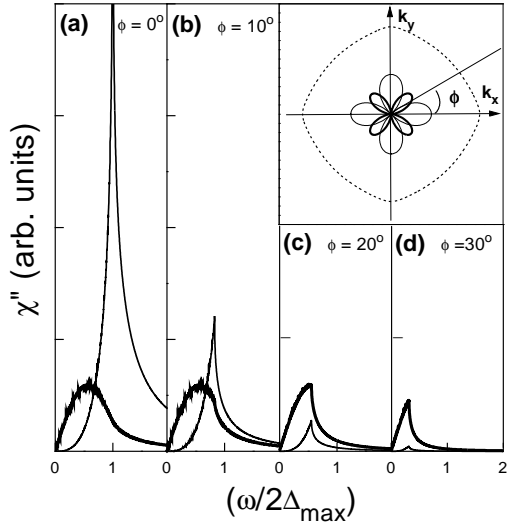}}
\vspace{0.14in}
\caption{Calculated B$_{1g}$ (thin lines) and B$_{2g}$ (thick lines)
Raman response functions in La214 (a) by integrating over the full Fermi 
surface (dashed lines in the inset) and (b-d) by integrating only over region
lying between $\phi$ and $90^{\circ} - \phi$ in each quadrant. The inset 
shows polar plots of the angular dependence of the 
B$_{1g}$ (thin lines) and B$_{2g}$ (thick lines) components of the 
Raman tensor $\gamma({\bf k})$.}
\label{5}
\end{figure}

To illustrate the dramatic change in the intensity in the B$_{1g}$ channel
and the relative constancy of the B$_{2g}$ channel, we can calculate the 
unscreened Raman response functions in the superconducting state
using the conventional model of light scattering \cite{klein}
\begin{equation}
\chi^{\prime\prime}_{\gamma}(\omega)  
\propto \, \langle\, \frac{\gamma^{2}({\bf k}) \Delta^{2}({\bf k})}
{\omega \sqrt{\omega^2-4\Delta^{2}({\bf k})}}\,\rangle,
\label{t0}
\end{equation}
where $\langle \cdots\rangle$ imply an average over the FS for k-vectors such 
that $\omega > 2|\Delta({\bf k})|$. The results shown in Fig. 5 were obtained 
with $\gamma({\bf k})$ calculated using a second neighbor tight binding 
model with the parameters described previously \cite{chen2} and 
$\Delta({\bf k})=\Delta_{0}[\cos(k_{x}a)-\cos(k_{y}a)]$. The response 
functions obtained by integrating (1) over the full FS (dashed lines in the 
inset of Fig. 5) are shown in Fig. 5a. To obtain a representation of 
the Raman response in the underdoped case we integrate (1) over portions of 
the FS lying between $\phi$ and $90^{\circ} - \phi$ in each quadrant.
That is, regions of the FS near the axis (defined by angle $\phi$) were
excluded from the integration, and the resulting response functions are
shown in Figs. 5b-5d for $\phi = \,$10$^{\circ}$, 20$^{\circ}$, and 
30$^{\circ}$. As is evident an increasing truncation of the FS leads to 
a rapid depletion of the B$_{1g}$ spectrum, but has much smaller  
effect on the B$_{2g}$ spectrum. 

We have also calculated the ratio R of the integrated response functions 
as a function of $\phi$ and the results are superimposed (dashed line)
on the data points in Fig. 4. For the purpose of this comparison
both the measured and calculated response functions were integrated
from $\omega = 0$ to 600$\,cm^{-1}$.
It should be noted that the apparent agreement between calculation and 
experiment is quite fortuitous given the simplistic nature of the model.
The results do, however, provide plausibility to the existence of 
localized depletion of spectral weight from regions of the FS located 
near ($\pi,0$). 
Finally we should note that the effective gapping of quasiparticles (QP) 
from these same region, at T $>$ T$_c$, is  consistent with the absence of 
a superconductivity induced renormalization in the B$_{1g}$ spectra of the 
underdoped compounds (Figs. 2a and 2b).

The observed dependence on scattering geometry thus implies that the spectral 
weight loss is confined to regions of the FS located near ($\pm \pi,0$) and 
($0,\pm \pi$) and that QP located near the diagonals are essentially 
unaffected by a reduction in doping level. 
Given that ARPES experiments \cite{marshall,harris,ding} also find a SWD that 
reaches a maximum in regions near ($\pi,0$), we will assume that this feature 
is characteristic of the PG, and that the PG thus manifests itself in the 
Raman spectra {\em primarily} as a SWD in the B$_{1g}$ channel. 
We should note however, that the B$_{2g}$ spectra is also 
approximately independent of temperature (for T$\,>\,{\rm T}_c$), which 
is difficult to reconcile with the FS contraction envisaged by Norman 
{\em et al.} \cite{norman2}.

According to Fig. 4 the SWD or pseudogap has approximately 
the same strength in La214 and Y123. This is not consistent with the 
suggestion \cite{mm} that the PG arises from strong AFM interactions between 
the closely spaced CuO$_2$ layers of bilayer compounds such as Y123.  
The present results thus suggest that the PG arises from intralayer 
interactions and in this regard are in agreement with IR 
\cite{puchkov,tatiana}, ARPES 
\cite{ino}, specific heat \cite{loram3}, and NMR \cite{yasuoka} measurements.  

\subsection{Magnitude (E$_g$) of the Pseudogap} 

In the spectra we have obtained from La214 there is no clear indication of  
an onset temperature T$^*$ for the PG. Estimates \cite{pines2,yasuoka} 
of T$^* \leq 200\,$K have been obtained from the analysis of NMR experiments 
carried out on slightly underdoped La214 samples (x = 0.14 and 0.15). 
On the other hand transport \cite{batlogg}, IR reflectivity \cite{tatiana},
and specific heat \cite{loram3} measurements on underdoped La214
have all found T$^*  > 300\,$K. 
Our spectra (Fig. 1) indicate some minor changes at about
200$\,$K, that include the appearance of a broad and weak feature at about
600$\,cm^{-1}$ in both channels. In fact the variations that occur in 
the B$_{2g}$ spectra below about 200$\,$K reflect a decrease in intensity
at low energies similar to that observed by Nemetschek  {\em et al}. 
\cite{hackl} in Bi2212. However, no significant changes in spectral weight, 
which one might expect to mark the onset of the PG, occur at any 
temperature below $300\,$K. 

The lack of an experimental consensus for T$^*$ in La214 is perhaps 
supportive of the suggestion \cite{tallon3} that one cannot associate a well 
defined T$^*$ with the PG, but only an energy scale E$_g$. In this regard, 
however, there is also no clear evidence of a gap edge in the B$_{1g}$ 
spectra we have obtained (Fig. 1) from underdoped La214. This is perhaps 
not surprising if E$_g$ is of order of J in underdoped samples and decreases 
with increasing hole concentration  as has been suggested by photoemission 
\cite{ino} and specific heat \cite{loram3} measurements. 
This magnitude for the PG might be expected if short range AFM 
correlation were responsible for the spectral weight depletion \cite{pines1}.
One possible scenario involves the existence of 
AFM ordered regions and separate hole rich regions in the underdoped 
samples. Evidence for this type of phase separation was obtained recently 
in the muon spin resonance experiments on underdoped La214 and Ca doped Y123
\cite{ndmyr}. 

\begin{figure}[htb]
\centerline{\epsfxsize=2.8in \epsffile{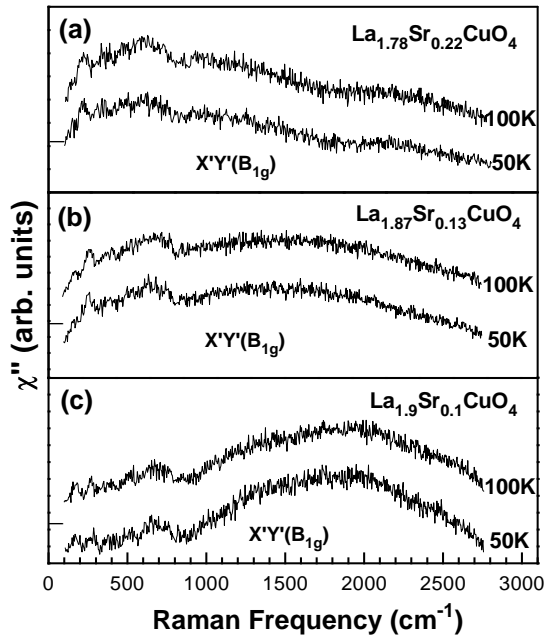}}
\vspace{0.12in}
\caption{The high energy B$_{1g}$ Raman response functions measured at
50$\,$K and 100$\,$K in La214(x) for (a) x = 0.22, (b) x = 0.13, and (c)
x = 0.10.
}
\label{6}
\end{figure}

In Raman spectra one-magnon excitations are not observed and hence the 
PG energy would be reflected in the two-magnon peak that occurs at about 
3J $\approx$ 3000$\,$cm$^{-1}$ in undoped La214 \cite{lyons,sugai} and 
insulating Y123 \cite{lyons,reznik,blum}. The amplitude and energy of the 
two-magnon peak decreases \cite{sugai,rubhausen} as the compounds are doped 
and such peaks have not been observed for doping levels greater than or equal 
to optimum. Although our spectrometer range is limited, 
we have carried out preliminary measurements of the high energy B$_{1g}$ 
spectra for three samples (x = 0.10, 0.13, and 0.22).
The high energy ($\omega \,<\, 3000\,cm^{-1}$) response functions measured 
at 50$\,$K and 100$\,$K for the three samples are shown in Fig. 6. 
From the figure one can see a broad, weak peak at about 2000$\,$cm$^{-1}$ 
(J$\,\sim 700\,$cm$^{-1}$) in the x = 0.10 spectrum. This peak 
essentially vanishes as we proceed to higher doping levels (Figs. 6a and 6b).
We must emphasize that these spectra have not been corrected for  
variations in the optical constants or for
spectrometer response and are shown only to illustrate the effect of doping.
Our spectrometer response decreases in the red and correction might
push the peak in Fig. 6c to higher energies and also decrease its relative
amplitude. Such corrections would not however alter the observed trend that is
induced by changes in doping.

It appears plausible that the pseudogap or SWD in the B$_{1g}$ 
channel is due to short range AFM correlations and the formation of 
``hot spots'' \cite{pines1} on regions of the FS located near ($\pm\pi,0$) 
and ($0,\pm\pi$).
As the doping level is increased the AFM interactions become weaker, 
as is evidenced by a weakening (Fig. 6) and shift of the two-magnon peak
to lower energies \cite{sugai,rubhausen}. Thus as the doping level is
increased, E$_g$ decreases and some spectral weight is shifted from higher 
to lower energies.  At optimum doping, where a SCIR is observed in the 
B$_{1g}$ channel (Fig. 3c), the PG is assumed to be filled. 
Furthermore on the basis of their specific heat measurements in several 
cuprates Loram {\em et al.} \cite{loram3} have proposed
that E$_g \approx\,$J$\,(1-\,p/p_{cr})$ where $p$ is the hole concentration 
and $p_{cr}$ the value at which E$_g$ ``closes''. Given that 
J$\,\approx 1000\,$cm$^{-1}$ in La$_2$CuO$_4$ and E$_g\,\approx\, 
700\,$cm$^{-1}$ (Fig. 6c) suggests $p_{cr}\, > \,p_{opt}$. We would  
thus conclude that the pseudogap ``fills'' before it ``closes''.

It is interesting to compare the PG energy scale E$_g$ 
to that of the superconducting gap $\Delta$. In Raman spectra the energy of 
the pair-breaking peak in the B$_{1g}$ channel can be used \cite{tom1}
as a measure of $\Delta$. In optimally doped La214 the maximum value of 
the superconducting gap \cite{chen2} is given by 
$\omega$(B$_{1g}) \approx 2\Delta_{max} \approx 200\,$cm$^{-1}$,
as is also evident from Fig. 3.
In the overdoped region $\Delta_{max}$ decreases (Fig. 3), consistent with 
the behavior in other cuprates \cite{chen3,white,kendziora}.
In the underdoped regime the absence of SCIR in the B$_{1g}$ channel
means that we do not have any direct measure of the gap maxima.
We can note however that the frequency of the pair-breaking peak in the 
B$_{2g}$ channel remains approximately constant 
[$\omega$(B$_{2g}) \approx 130 \pm 20\,$cm$^{-1}$] as the doping level is 
decreased from optimum.  We thus infer that $\Delta$ remains approximately 
constant throughout the doping range $0.10 \leq {\rm x} \leq 0.17$.
This behavior is also consistent with that found in ARPES and tunneling
measurements \cite{harris,renner} on Bi2212. Therefore we have an amplitude
$\Delta_{max} \approx 100\,$cm$^{-1}$ for the SC gap and E$_g \sim\,$J$\,\sim
700\,$cm$^{-1}$ for the PG. These very different energy scales suggest 
that different physical mechanisms are involved.
\subsection{Summary}

The most noticeable aspect of the spectra shown in Fig. 3 is the 
SWD that occurs in the B$_{1g}$ channel when the doping level is reduced 
below optimum, while the strength of the B$_{2g}$ spectrum is essentially
unaffected by changes in doping. 
These results  suggest that the QP scattering rate is highly anisotropic 
\cite{tom2} and leads to  the existence of ``hot spots'' 
on regions of the FS located near ($\pm\pi,0$) and ($0,\pm\pi$). 
This observation is consistent with the predictions of Pines and 
coworkers \cite{pines1} who have pointed out that QP located near these 
regions can be coupled by 
the AFM ordering vector {\bf Q} = ($\pi,\pi$) and hence interact strongly
in underdoped compounds. Conversely, the QP on regions of the FS located 
near  $|{\bf k}_x|=|{\bf k}_y|$ are weakly coupled
\cite{pines1}, and have been designated as ``cold''  \cite{pines1,ioffe}. 
As the doping level is increased toward optimum the SWD decreases rapidly, 
consistent with a weakening of the AFM interactions between QP \cite{pines1}.  
In this language our results then suggest that the hot QP are gapped above
T$_c$ and hence are not affected by the SC transition. Only the cold QP,
as evidenced by the SCIR in the B$_{2g}$ channel at T$_c$, 
participate in the transition to the SC state.

Thus far our attention has been focused on the most dominant feature of the
spectra obtained from underdoped compounds, namely the SWD that is confined 
to the B$_{1g}$ channel. As mentioned previously Nemetschek {\em et al.}
\cite{hackl} found that a relatively small depletion of spectral weight
also occurs in the B$_{2g}$ channel of Y123 and Bi2212 below
$\omega \approx 700\,$cm$^{-1}$ and T$\approx 250\,$K. This energy and onset
temperature are in good agreement with IR measurements on the same compounds
\cite{puchkov}. This appears reasonable since, given the SWD in 
the B$_{1g}$ channel one would expect the results of IR and transport 
measurements, for example, to be determined mainly by the properties of the 
cold QP \cite{hackl,pines1,ioffe}, or those on regions of the FS located near 
the diagonal directions. Furthermore, one can see that a similar small 
depression of spectral weight appears below 700$\,$cm$^{-1}$ and 200$\,$K 
in the B$_{2g}$ spectra of the x = 0.13 crystal (Fig. 1b), in agreement with 
IR measurements carried out \cite{tatiana} on underdoped La214.
Thus the loss of spectral weight in the B$_{2g}$ channel \cite{hackl}, 
the depression of the scattering rate observed in IR experiments, and 
the spectral weight depletion observed in the B$_{1g}$ channel (Figs. 3 and 4)
are all characterized  by the same energy scale. This suggests that there
is a common physical mechanism underlying the apparently different phenomena
that are observed in the two channels. 
However, the identification of a possible interrelation between the 
two channels will require a very careful analysis of 
temperature dependence of the spectra for 50$\,$K$\leq\,$T$\,\leq 300\,$K. 
Such an analysis should also provide important information about the existence 
and magnitude of crossover temperatures T$^*$ that have
been observed in other experiments \cite{tatiana,pines2,yasuoka,batlogg}
and are important features of some models \cite{emery,pines1}.
It is also possible that the SWD in the two channels arises from different
mechanisms and that there are different pseudogaps as have been suggested
 \cite{ino} by  Ino {\em et al}.

\section{Conclusions}
In the Raman spectra obtained from La214 there is a loss of low energy 
spectral weight in the B$_{1g}$ channel as the doping level is decreased 
below optimum, while the intensity in the B$_{2g}$ channel remains unchanged.
These results imply the existence of a strong anisotropic 
scattering mechanism and the formation of hot spots \cite{pines1} on regions 
of the Fermi surface located near k$_x$ and k$_y$ axes. 
Furthermore, in underdoped compounds the B$_{1g}$ channel is unaffected by 
the superconducting transition.
The pseudogap is thus manifested in Raman spectra both by a depletion 
of low energy spectral weight or reduced intensity and by the absence of a 
superconductivity induced renormalization in the B$_{1g}$ channel.
In this context the pseudogap in La214 and Y123 \cite{chen1} is present 
only in the underdoped regime. Since the spectral weight depletion
is similar in the two compounds, we conclude that the 
pseudogap is intrinsic to a single CuO$_2$ plane and does not arise
from interlayer interactions.
Furthermore the results suggest that the loss of low energy  
intensity in the B$_{1g}$ channel is due to strong antiferromagnetic 
interactions which in turn implies E$_g \approx\,$J$\,\approx 700\,$cm$^{-1}$ 
for the pseudogap, an energy scale that is much larger than the energy  
of the superconducting gap ($\Delta_{max} \approx 100\,$cm$^{-1}$).
These energy scales imply that the pseudogap and superconducting gap 
arise from different physical mechanisms. Since superconductivity grows 
while the spectral weight loss or  pseudogap diminishes, it appears that 
the two mechanisms compete with each other for the available quasiparticles.

\section{Acknowledgements}
The authors gratefully acknowledge useful conversations with T. P. Devereaux, 
T. Startseva, T. Timusk, and J. Schmalian. One of us (JCI) has also benefited greatly
from discussions with R. G. Buckley, J. L. Tallon, A. J. Trodhal, and 
G. V. M. Williams during a visit to the Industrial Research Laboratory in 
Lower Hutt, New Zealand. The financial support of the Natural Sciences and 
Engineering Research Council of Canada is gratefully acknowledged.

{\small {$^{\ast}$Present address: Department of Material 
Science, Shizuoka University, Johoku, 
Hamamatsu 432, Japan.}}\\
{\small {$^{\dagger}$Present address: Joint Research Center for 
Atom Technology, Higashi, Tsukuba 305, Japan.}}

\end{document}